\documentclass[aps,prd,eprint,numerical,showpacs,showkeys,10pt]{revtex4-2}

\usepackage{longtable}
\usepackage{amsmath}
\usepackage{amssymb}
\usepackage{amsfonts}
\usepackage{graphicx}
\usepackage{dcolumn}
\usepackage{bm}
\usepackage{hyperref}
\usepackage{color}
\hypersetup{colorlinks=false}

\begin{document}

\title{Extending the model of rotating acoustic geometries to include nonvanishing solid-body rotation: Quasibound spectra}

\author{H. S. Vieira}
\email{horacio.santana.vieira@hotmail.com or horacio.santana.vieira@ifsc.usp.br}
\affiliation{Sao Carlos Institute of Physics, University of Sao Paulo, 13566-590, Sao Carlos, SP, Brazil}

\date{\today}

\begin{abstract}
In a very recent paper, we computed the quasibound states of massless acoustic excitations interacting with a new (effective) acoustic geometry with circulation. Notably, the behavior of this acoustic black hole aligns with the phenomenology observed in recent experiments that include superfluids. Such vortex fluid flow bundles exhibit solid-body rotation at length scales larger than the inter-vortex distance, which adds some complexity to the study of quantum fluid behavior. To theoretically deal with this issue, we present an extension of our previous results by including a solid-body rotation term at the angular fluid velocity and then we perform a spectral study by using the analytical solutions for the scalar wave equation of motion. The resulting analytically solvable framework bridges idealized acoustic black-hole models and experimentally motivated rotating-fluid configurations, providing a more realistic description of quasibound state spectra in analog gravity.
\end{abstract}


\maketitle

%
%%%%%%%%%%%%%%%%%%%%%%%%%%%%%%%%%%%%%%%%% Introduction
%
\section{Introduction}\label{Introduction}
%
%%%%%%%%%%%%%%%%%%%%%%%%%%%%%%%%%%%%%%%%%
%
In rotating fluid experiments, it is well established that the equilibrium depth of a fluid in solid-body rotation depends on the rotation axis, the angular velocity, the gravitational acceleration, and the fluid depth at the rotation axis. Under these conditions, the free surface assumes a parabolic profile, adjusting such that it remains perpendicular to the effective gravity vector, defined as the combination of gravitational and centrifugal accelerations. A fluid contained in a rotating vessel and isolated from external mechanical and thermodynamic forcing will, after a transient adjustment, reach a solid-body rotation state in which there is no relative motion between the fluid and the container. In this equilibrium configuration, the free surface is therefore nonplanar and adopts a paraboloidal shape, consistent with the balance of forces acting on the fluid \cite{PhysRevD.110.104058,PhysRevD.110.124068,PhysRevD.111.104064,MIT:weathertank}.

This equilibrium configuration provides a useful reference state for rotating-flow experiments and theoretical analyses. The parabolic free-surface profile constitutes a direct consequence of rotational dynamics and serves as a diagnostic of the effective potential governing the system. As such, it offers a controlled framework for investigating deviations from solid-body rotation, as well as for modeling rotational effects in analogous physical systems where centrifugal and gravitational forces play a central role \cite{PhysRevD.96.064014,EurPhysJC.86.227}.

Solid-body rotation refers to the equilibrium state in which every fluid element rotates with the same constant angular velocity $\Omega$ about a common axis, so that the azimuthal velocity field is given by $v_\phi(r)=\Omega r$. In contrast to the irrotational circulation typically employed in acoustic black hole models, whose azimuthal velocity scales as $v_\phi\propto 1/r$ outside the vortex core, solid-body rotation possesses a uniform vorticity, $\nabla\times\mathbf{v}=2\Omega\,\hat{\mathbf{z}}$, and therefore represents a fundamentally different flow configuration. In realistic laboratory experiments, however, these two contributions need not be mutually exclusive. Residual solid-body rotation may persist owing to imperfect cancellation of the global angular momentum imparted by the experimental apparatus, producing a background rotational flow superimposed on the circulating fluid. Although this contribution is generally small, it modifies the equilibrium background and, consequently, the effective acoustic metric experienced by linear perturbations. Quantifying its influence is therefore essential for assessing the extent to which theoretical predictions based on idealized irrotational backgrounds remain applicable to experimentally realizable analog black hole systems.

In superfluid experiments, residual solid-body rotation (with angular frequency $\Omega$) constitutes an undesired component that can leak into the experimental region through the flow-conditioning system. To mitigate this effect, the experimental setup is designed such that the propeller is driven in the direction opposite to the orientation of the apertures through which the fluid enters the chamber. This procedure leads to a sparse distribution of quantized vortices, thereby partially justifying the assumption that the normal and superfluid components may be treated as a single effective fluid. Experimental measurements indicate that the ratio between $\Omega$ and the angular driving frequency does not exceed 2.5\% \cite{Nature.628.66}. Nevertheless, a finite degree of solid-body rotation remains present in the system.

The persistence of this residual rotation is of particular relevance for the interpretation of experimental results, as even a small background angular velocity may influence the large-scale flow structure and the dynamics of quantized vortices. Consequently, the assumption of strictly irrotational flow must be regarded as an approximation, whose validity depends on the sensitivity of the observables under consideration. Accounting for the effects of residual solid-body rotation is therefore essential when comparing experimental data with theoretical models that rely on idealized rotational states \cite{PhysRevD.110.024035}.

Although such experiments generate quantized vortices rather than acoustic black holes, they provide a foundational platform for future investigations in analog gravity. In principle, these setups can be straightforwardly extended by introducing a singularity in the flow sink, thereby reproducing the effective geometry of a rotating acoustic black hole. It is therefore of interest to examine the role played by residual solid-body rotation in shaping the spectrum of quasibound states (QBSs).

From a theoretical perspective, the presence of solid-body rotation modifies the effective background flow and, consequently, the associated acoustic metric. Such modifications may lead to measurable shifts in the QBS spectrum, potentially affecting mode localization and stability properties. A systematic analysis of these effects is thus necessary in order to disentangle genuine features of the analog black hole geometry from artifacts arising from background rotation.

In a recent study \cite{PhysRevD.111.104025}, we employed Unruh's acoustic metric for fluid flows, incorporating circulation through a nonvanishing angular velocity component in order to construct an analog of a rotating black hole in the vicinity of an acoustic event horizon defined by the radial flow velocity. The linear response of this system to small perturbations was then analyzed, and the corresponding excitation spectrum was obtained. In the present work, we focus on novel features of the spinning acoustic metric arising from the inclusion of a solid-body rotation term in the angular component of the background flow. In particular, we examine the imprint of this additional contribution on the spectrum of small excitations. In the following Sections, we compute the QBSs associated with this effective geometry and present the results across the full parameter space of the metric.

Unlike the idealized irrotational models commonly considered in analog gravity studies, the present construction incorporates a residual solid-body rotation motivated by recent laboratory observations. Although the residual solid-body rotation is experimentally small, its inclusion changes the mathematical structure of the effective background by introducing finite vorticity, thereby taking the model beyond the class of globally irrotational acoustic geometries. This modification enables us to investigate, within an analytically solvable framework, how experimentally motivated departures from ideal irrotationality affect the QBS spectrum. In this sense, the present work is not merely a quantitative correction to previous models, but a step toward bridging the gap between idealized analog spacetimes and realistic laboratory configurations.

This analysis extends previous investigations of analog black hole spectroscopy by accounting for background rotational effects that are expected to be present in realistic experimental configurations. By systematically characterizing the influence of solid-body rotation on the QBS spectrum, our results provide a more comprehensive theoretical framework for interpreting experimental data. In conjunction with recent advances in analog gravity experiments, this study may therefore contribute to the identification and detection of quasibound states in laboratory realizations of rotating acoustic black holes. Despite the additional complexity introduced by the residual solid-body rotation, the model remains analytically tractable, allowing the complete QBS spectrum to be obtained in closed form through confluent Heun functions.

Since QBS frequencies are determined by the detailed structure of the effective potential, they constitute particularly sensitive probes of small modifications in the background geometry. Consequently, even weak residual solid-body rotation may produce measurable spectral shifts.

The paper is organized as follows. In Sec. \ref{UABHc} we present the Unruh acoustic black hole metric with circulation and, then, we add the solid-body rotation in Sec. \ref{UABHcr}. In Sec. \ref{WE} we solve the massless Klein-Gordon equation in the background under consideration. In Sec. \ref{QBSs} we present the spectrum of QBSs. Finally, in Sec. \ref{Conclusions} we give our final remarks.
%
%%%%%%%%%%%%%%%%%%%%%%%%%%%%%%%%%%%%%%%%% Unruh acoustic black holes with circulation: historical motivation
%
\section{Unruh acoustic black holes with circulation: historical motivation}
\label{UABHc}
%
%%%%%%%%%%%%%%%%%%%%%%%%%%%%%%%%%%%%%%%%%
%
The purpose of this Section is solely to summarize the standard derivation of Unruh'sacoustic metric, which serves as the historical and conceptual starting point of the present work. The phenomenological extension introduced in the following Section is not derived from these equations, but is instead motivated by experimentally observed residual solid-body rotation.

We begin by considering a general acoustic black hole geometry embedded in Minkowski spacetime, as originally formulated by Unruh~\cite{PhysRevLett.46.1351}. Particular attention is given to the choice of the radial component of the fluid velocity profile. The fundamental equations governing the dynamics of an irrotational fluid are given by
\begin{eqnarray}
\nabla \times \mathbf{v} & = & 0, \label{eq:irrotational}\\
\partial_{t}\rho+\nabla\cdot(\rho\mathbf{v}) & = & 0, \label{eq:Continuity}\\
\rho[\partial_{t}\mathbf{v}+(\mathbf{v}\cdot\nabla)\mathbf{v}] +\nabla p & = & 0. \label{eq:Euler}
\end{eqnarray}
Here, $\mathbf{v}$, $\rho$, and $p$ are the velocity, density, and pressure of the fluid, respectively. We now introduce the velocity potential $\Psi$, defined such that $\mathbf{v} = -\nabla \Psi$, and assume the fluid to be barotropic, i.e., $\rho = \rho(p)$. We then proceed by linearizing the equations of motion around a given background configuration $(\rho_{0}, p_{0}, \Psi_{0})$, namely,
\begin{eqnarray}
\rho & = & \rho_{0}+\epsilon\rho_{1}, \label{eq:rho}\\
p & = & p_{0}+\epsilon p_{1}, \label{eq:p}\\
\Psi & = & \Psi_{0}+\epsilon\Psi_{1}, \label{eq:psi}
\label{eq2:Madelung_representation}
\end{eqnarray}
to obtain the following wave equation:
\begin{equation}
-\partial_{t}\biggl[c_s^{-2}\rho_{0}(\partial_{t}\Psi_{1}+\mathbf{v}_{0}\cdot\nabla\Psi_{1})\biggr]+\nabla\cdot\biggl[\rho_{0}\nabla\Psi_{1}-c_s^{-2} \rho_{0}\mathbf{v}_{0}(\partial_{t}\Psi_{1}+\mathbf{v}_{0}\cdot\nabla\Psi_{1})\biggr]=0.
\label{eq:wave_equation_Visser}
\end{equation}
Here, $c_{s}$ is the local speed of sound defined by
\begin{equation}
c_{s}^{-2} \equiv \frac{\partial \rho}{\partial p}.
\label{eq:sound}
\end{equation}
The wave equation (\ref{eq:wave_equation_Visser}) governs the propagation of the linearized scalar potential $\Psi_{1}$, i.e., it describes the dynamics of phase fluctuations treated as small perturbations in a homogeneous and stationary fluid. This equation can be recast in the form of a wave equation in an effective curved spacetime,
\begin{equation}
\frac{1}{\sqrt{-g}}\partial_{\mu}\left(g^{\mu\nu}\sqrt{-g}\,\partial_{\nu}\Psi_{1}\right)=0,
\label{eq:KG_equation}
\end{equation}
which is formally equivalent to the covariant massless Klein-Gordon equation. The associated acoustic line element is given by
\begin{equation}
ds^{2} = \frac{\rho_{0}}{c_{s}}\biggl[-c_{s}^{2}\,dt^{2}+(dx^{i}-v_{0}^{i}\,dt)\delta_{ij}(dx^{j}-v_{0}^{j}\,dt)\biggr].
\label{eq:acoustic_metric}
\end{equation}
In the original construction, Unruh assumed a spherically symmetric, stationary, and inviscid background flow. Now, by incorporating circulation -- namely, a rotational contribution -- into the radial velocity profile, the acoustic metric in Eq.~(\ref{eq:acoustic_metric}) can be generalized to
\begin{equation}
ds^{2} = \frac{\rho_{0}}{c_{s}}\biggl\{-\biggl[c_{s}^{2}-(v_{0}^{r})^{2}-(v_{0}^{\phi})^{2}\biggr]\,dt^{2}+\frac{c_{s}}{c_{s}^{2}-(v_{0}^{r})^{2}}\,dr^{2}-2v_{0}^{\phi}r\sin\theta\,dt\,d\phi+r^{2}(d\theta^{2}+\sin^{2}\theta\,d\phi^{2})\biggr\},
\label{eq:Unruh_circulation}
\end{equation}
where the following coordinate transformations have been implemented:
\begin{eqnarray}
&& dt \rightarrow dt-\frac{v_{0}^{r}}{c_{s}^{2}-(v_{0}^{r})^{2}}\,dr,\\
&& d\phi \rightarrow d\phi-\frac{v_{0}^{\phi}v_{0}^{r}}{r\sin\theta[c_{s}^{2}-(v_{0}^{r})^{2}]}\,dr,
\label{eq:coordinate_transformations}
\end{eqnarray}
with the velocity field given by $\mathbf{v}=v_{0}^{r}\mathbf{e}_{r}+v_{0}^{\phi}\mathbf{e}_{\phi}$. Owing to the rotational character of the geometry, the condition $c_{s}^{2}-(v_{0}^{r})^{2}-(v_{0}^{\phi})^{2}=0$ defines the location of the ergoregion. This surface plays a central role in processes associated with energy extraction from rotating black holes, and is directly related to the occurrence of superradiant scattering \cite{JETP.35.1085,NaturePhys.13.833,LectNotesPhys.971}. In this context, the presence of an ergoregion in the acoustic spacetime suggests that analogous superradiant effects may arise for sound waves propagating in the fluid. In particular, incident modes with appropriate frequency and angular momentum can be amplified upon scattering off the rotating background, extracting energy from the flow. This provides a promising framework for the experimental investigation of superradiance in analog gravity systems, where controlled laboratory setups may reproduce key features of rotating black hole spacetimes.

If the background radial flow smoothly transitions to the supersonic regime at $r = r_{h}$, then $r_{h}$ defines an acoustic horizon. In the vicinity of this surface, the radial component of the velocity field can be expanded as
\begin{equation}
v_{0}^{r} = -c_{s} + a(r - r_{h}) + \mathcal{O}((r - r_{h})^{2}),
\label{eq:radial_velocity}
\end{equation}
where the parameter $a$ is defined by
\begin{equation}
a = (\nabla \cdot \mathbf{v})\big|_{r = r_{h}},
\label{eq:a_4DUABH}
\end{equation}
and carries dimensions of frequency. The angular component of the flow velocity can be expressed as
\begin{equation}
v_{0}^{\phi} = \frac{J}{r} = \frac{C \sin\theta}{r},
\label{eq:angular_velocity}
\end{equation}
where $J$ corresponds to the angular momentum of the fluid, while $C$ denotes the circulation. That is the background spacetime presented in Ref.~\cite{PhysRevD.111.104025}.

Next, we are going to include a residual solid-body rotation in this background flow, designed to mimic the general azimuthal vortex motion generated by an external experimental setup. By adopting an appropriate ansatz for the angular component of the velocity field, we achieve an effective acoustic metric that approaches that of a rotating black hole in the vicinity of the acoustic horizon.
%
%%%%%%%%%%%%%%%%%%%%%%%%%%%%%%%%%%%%%%%%% Effective acoustic black holes with circulation and solid-body rotation
%
\section{Effective acoustic black holes with circulation and solid-body rotation}
\label{UABHcr}
%
%%%%%%%%%%%%%%%%%%%%%%%%%%%%%%%%%%%%%%%%%
%
The derivation summarized in Sec. \ref{UABHc} corresponds to the standard irrotational acoustic metric introduced by Unruh. In the present work, however, we no longer interpret the resulting scalar equation as arising from perturbations of a globally defined velocity potential. Instead, Eq.~(\ref{eq:KG_equation}) is adopted as an effective scalar wave equation propagating on a phenomenological acoustic geometry motivated by recent rotating-superfluid experiments.

We consider a phenomenological extension of Unruh's acoustic metric aimed at modeling laboratory configurations in which a residual solid-body rotation coexists with a circulating vortex. Our purpose is not to derive this geometry from the equations of ideal hydrodynamics, but rather to investigate the propagation of effective test scalar fields on a background motivated by experimentally observed flows.

The angular component of the flow velocity can be now expressed as
\begin{equation}
v_{0}^{\phi} = \frac{J}{r} + \Gamma\,r = \frac{C \sin\theta}{r} + \Omega\sin\theta\,r,
\label{eq:angular_velocity_cr}
\end{equation}
where $\Gamma$ corresponds to the solid-body rotation of the fluid, while $\Omega$ denotes the associated angular frequency. It is worth emphasizing that the solid-body rotation considered in this work is intended as an effective description of the finite experimental region rather than a global flow extending to arbitrarily large distances. Since the azimuthal velocity associated with solid-body rotation increases linearly with radius, such a profile cannot remain physically valid at sufficiently large radii, where the assumptions underlying the acoustic metric would eventually break down. In realistic laboratory configurations, however, the fluid is confined within a finite container, and the background flow is expected to undergo a transition near the boundaries owing to geometric constraints and the driving mechanism. Consequently, our analysis focuses on the region where the effective acoustic geometry is well defined and where the quasibound states are localized. In this sense, the present model should be regarded as a local approximation describing the experimentally relevant portion of the flow, rather than an asymptotically extended spacetime analogous to astrophysical black hole geometries. With these definitions, the acoustic line element describing a rotating black hole with circulation and solid-body rotation can be written as
\begin{equation}
ds^{2} = -\biggl[f(r) - \frac{J^{2}}{r^{2}} - \Gamma^{2}r^{2} - 2J\Gamma \biggr]\,dt^{2} + \frac{1}{f(r)}\,dr^{2} - 2(J+\Gamma\,r^{2})\sin\theta\,dt\,d\phi + r^{2}(d\theta^{2} + \sin^{2}\theta\,d\phi^{2}),
\label{eq:metric_UABHc}
\end{equation}
where the metric function is given by
\begin{equation}
f(r) = 2a(r - r_{h}).
\label{eq:metric_function_UABHc}
\end{equation}
Here, $r_{h}$ denotes the acoustic event horizon, defined as the outermost marginally trapped surface for outgoing phonons, at which the radial flow becomes supersonic. For simplicity, we have set $c_{s} = 1$ and assumed a constant background density $\rho_{0}$, thereby omitting an overall position-independent conformal factor. In addition, the acoustic surface gravity -- interpreted as the effective gravitational acceleration at the acoustic horizon -- is defined by
\begin{equation}
\kappa_{h} \equiv \frac{1}{2}\frac{df(r)}{dr}\biggr|_{r = r_{h}} = a,
\label{eq:surface_gravity_UABHc}
\end{equation}
where the last equality follows directly from the form of the metric function $f(r)$. Moreover, the angular velocity associated with frame dragging in this acoustic geometry is given by
\begin{equation}
\varpi(r) \equiv -\frac{g_{0\phi}}{g_{\phi\phi}} = \frac{C}{r^{2}} + \Omega,
\label{eq:dragging_UABHc}
\end{equation}
so that its value at the acoustic horizon, $\varpi_{h} \equiv \varpi(r_{h}) = \frac{C}{r_{h}^{2}} + \Omega$, characterizes the dragging of inertial frames at $r = r_{h}$.

The effective metric (\ref{eq:metric_UABHc}) is regarded as a phenomenological analog spacetime. The scalar field $\Psi_{1}$ is no longer interpreted as the perturbation of a global velocity potential, but simply as an effective test scalar field propagating on this background. In fact, the present geometry should not be regarded as an exact hydrodynamical acoustic metric, but as an effective spacetime capturing the dominant kinematical influence of a residual experimental rotation.

The new geometry described by Eq.~\eqref{eq:metric_UABHc} always admits an acoustic horizon located at $r = r_{h}$, and thus constitutes an analog model of a simplified rotating black hole. Experimental realizations reported in the literature -- such as the setup discussed in Ref.~\cite{Nature.628.66} -- can, however, exhibit horizonless configurations. In particular, two distinct regimes were identified: one without the formation of a depthless hollow, and another characterized by the emergence of a depthless hollow vortex. The latter configuration more closely resembles black hole spacetimes. In this sense, the effective metric in Eq.~\eqref{eq:metric_UABHc} provides a useful description of such (super)fluid vortex geometries at the kinematical level. It captures the essential features of a rotating analogue spacetime, albeit in the absence of nonlinear dispersive effects, which are present in the experimental system considered in Ref.~\cite{Nature.628.66}.

In what follows, we investigate the dynamics of effective test scalar fields propagating in the exterior region of the acoustic vortex spacetime described by Eq.~\eqref{eq:metric_UABHc}, and compute their quasibound spectrum. To this end, we employ the Vieira-Bezerra-Kokkotas (VBK) approach to determine the QBSs, which correspond to modes localized within the effective potential well of the acoustic geometry and decaying asymptotically at spatial infinity. Within this framework, the equation of motion is reduced -- via an appropriate separation of variables -- to a radial equation of Heun-like form.

It is important to emphasize that the introduction of the residual solid-body rotation takes the background flow beyond the class of globally irrotational configurations for which Unruh's acoustic metric is rigorously derived. Therefore, in the present work the geometry including the solid-body contribution should be understood as an effective phenomenological extension motivated by the experimental background flow. Consequently, the field $\Psi_{1}$ is interpreted as an effective test scalar field propagating on this modified acoustic geometry, rather than as the perturbation of a globally defined velocity potential. Our goal is to investigate how this experimentally motivated correction affects the quasibound state spectrum, while remaining within the standard analog gravity kinematical framework.

In what follows, motivated by the standard analogue-gravity framework, we consider the propagation of an effective test scalar field satisfying the covariant Klein-Gordon equation on the modified acoustic geometry.
%
%%%%%%%%%%%%%%%%%%%%%%%%%%%%%%%%%%%%%%%%% Limitations of the effective description
%
\subsection{Limitations of the effective description}
\label{Limitations}
%
%%%%%%%%%%%%%%%%%%%%%%%%%%%%%%%%%%%%%%%%%
%
The effective geometry introduced in this work is intended as a phenomenological extension of the standard acoustic metric rather than as an exact solution of the equations governing ideal hydrodynamics. As such, several assumptions and limitations should be emphasized.

First, the inclusion of the solid-body rotation component is motivated by recent laboratory observations of rotating superfluids, where a small residual background rotation may coexist with the circulating vortex. Consequently, the velocity profile considered here is not globally irrotational, and the effective metric should therefore not be interpreted as the exact acoustic metric obtained from the conventional linearization procedure of an inviscid, barotropic fluid.

Second, the present model is local in nature. The solid-body rotation is assumed to represent the experimentally relevant region surrounding the acoustic horizon, where the effective geometry provides a suitable approximation to the background flow. The model is not intended to describe the asymptotic behavior of the fluid at arbitrarily large distances, where additional physical effects and boundary conditions become important.

Third, throughout this work the field $\Psi_1$ is interpreted as an effective test scalar field propagating on the phenomenological acoustic geometry introduced in Eq.(\ref{eq:metric_UABHc}). It is not identified with the perturbation of a globally defined velocity potential. Accordingly, the covariant Klein-Gordon equation is adopted as the effective equation governing wave propagation on this background, in the same spirit as test-field analyses commonly employed in curved spacetime.

Finally, the purpose of the present construction is not to reproduce the full hydrodynamic dynamics of rotating superfluids, but rather to isolate and quantify the influence of experimentally motivated residual rotation on the quasibound state spectra. Within this effective framework, the resulting analytical solutions provide a useful theoretical laboratory for exploring how small departures from the ideal irrotational configuration modify the spectrum of acoustic excitations and may guide future analog-gravity experiments.

These assumptions are deliberately adopted in order to isolate the kinematical consequences of residual solid-body rotation while preserving an analytically solvable model.
%
%%%%%%%%%%%%%%%%%%%%%%%%%%%%%%%%%%%%%%%%% Scalar wave equation of motion
%
\section{Scalar wave equation of motion}\label{WE}
%
%%%%%%%%%%%%%%%%%%%%%%%%%%%%%%%%%%%%%%%%%
%
We now focus on the main properties of the effective geometry described by Eq.~\eqref{eq:metric_UABHc}, with particular emphasis on the propagation of effective test scalar fields and the associated QBS spectra. To this end, we solve the wave equation \eqref{eq:KG_equation} in this background.

Although the metric \eqref{eq:metric_UABHc} is not exactly spherically symmetric -- and therefore does not possess an exact separability property analogous to that of the Kerr spacetime -- it differs only perturbatively from the spherically symmetric acoustic geometry. Motivated by this weak departure, and consistently with the slowly rotating approximation adopted throughout this work, we employ the following separable ansatz as a leading-order approximation for the effective test scalar field:
\begin{equation}
\Psi_{1}(t,r,\theta,\phi) \simeq e^{-i \omega t} U(r) P_{\nu}^{m}(\cos\theta) e^{i m \phi},
\label{eq:ansatz_UABHc}
\end{equation}
where $\omega$ denotes the frequency, $U(r)=R(r)/r$ is the radial function, and $P_{\nu}^{m}(\cos\theta)$ are the associated Legendre functions with degree $\nu \in \mathbb{C}$ and order $m \in \mathbb{Z}_{+}$. The parameters $\nu$ and $m$ play the role of angular and azimuthal quantum numbers, respectively. In the context of QBSs, $\nu$ is, in general, complex, as will be discussed below. Higher-order corrections arising from the weak breaking of spherical symmetry are beyond the scope of the present work and are expected to modify the spectrum only perturbatively. Thus, by substituting Eqs.~(\ref{eq:angular_velocity})-(\ref{eq:ansatz_UABHc}) into the wave equation \eqref{eq:KG_equation}, we obtain the radial master equation
\begin{equation}
R''(r)+\frac{f'(r)}{f(r)}R'(r)+\frac{r^{2}[\omega-m\varpi(r)]^{2}-f(r)[\lambda+rf'(r)]}{r^{2}f^{2}(r)}R(r)=0,
\label{eq:radial_equation_UABHc}
\end{equation}
or, equivalently,
\begin{equation}
R''(r)+\frac{1}{r-r_{h}}R'(r)+\frac{[r^{2}(\omega-m\Omega)-Cm]^{2}-2a r^{2}(r-r_{h})(\lambda+2ar)}{4a^{2}r^{4}(r-r_{h})^{2}}R(r)=0,
\label{eq:radial_equation_UABHc_2}
\end{equation}
where $\lambda = \nu(\nu+1)$ is the separation constant. From Eq.~\eqref{eq:radial_equation_UABHc}, it follows that the condition for superradiant amplification is $\omega < \omega_{c}(r)$, where the critical frequency is given by $\omega_{c}(r) = m\varpi(r)$.
%
%%%%%%%%%%%%%%%%%%%%%%%%%%%%%%%%%%%%%%%%% Quasibound states
%
\section{Quasibound states}\label{QBSs}
%
%%%%%%%%%%%%%%%%%%%%%%%%%%%%%%%%%%%%%%%%%
%
In this Section, we apply the VBK approach to transform the radial equation (\ref{eq:radial_equation_UABHc_2}) into a confluent Heun equation, without imposing any boundary conditions \textit{a priori}. A detailed discussion of the VBK approach can be found in Refs.~\cite{PhysRevD.105.045015,EurPhysJC.82.669,EurPhysJC.82.932}. Once the exact analytical radial solution is obtained, we impose physically motivated boundary conditions, namely, purely ingoing waves at the acoustic horizon and exponentially decaying behavior at infinity. This procedure defines an eigenvalue problem for complex frequencies, whose solutions correspond to the QBS spectrum of the rotating vortex effective geometry. These complex frequencies characterize the quasibound states: their real part determines the oscillation frequency, while the imaginary part encodes the decay rate of the modes.

The VBK approach allows one to recast the radial equation (\ref{eq:radial_equation_UABHc_2}) into the confluent Heun form
\begin{equation}
y''(x)+\biggl(\alpha+\frac{1+\beta}{x}+\frac{1+\gamma}{x-1}\biggr)y'(x)+\biggl(\frac{\xi}{x}+\frac{\zeta}{x-1}\biggr)y(x)=0,
\label{eq:CHE_UABHc}
\end{equation}
where the dimensionless radial coordinate is defined as $x=1-r_h/r$, as well as the new radial wave function as $y(x)=x^{-\beta/2}(x-1)^{-\gamma/2}e^{-x\alpha/2}R(x)=\mbox{HeunC}(\alpha,\beta,\gamma,\delta,\eta;x)$. The coefficients $\alpha$, $\beta$, $\gamma$, $\delta$, and $\eta$ depend on the parameters $\omega$, $m$, $a$, $r_{h}$, $C$, and $\Omega$, and are explicitly given by
\begin{eqnarray}
\alpha	& = & \frac{im\varpi_{h}}{a},\label{eq:alpha_UABHc}\\
\beta	& = & -\frac{i(\omega-m\varpi_{h})}{a},\label{eq:beta_UABHc}\\
\gamma	& = & \frac{\sqrt{4a^{2}-(\omega-m\Omega)^{2})}}{a},\label{eq:gamma_UABHc}\\
\delta	& = & -\frac{m^2 C^2}{2 a^2 r_h^4},\label{eq:delta_UABHc}\\
\eta		& = & \frac{m^2 C^2 + r_h^3 [2 a^2 r_h + a \lambda - r_h (\omega - m \Omega)^2]}{2 a^2 r_h^4}.
\label{eq:eta_UABHc}
\end{eqnarray}
Here, the parameters $\xi$ and $\zeta$ are given by the following expressions
\begin{eqnarray}
\xi		& = & \frac{1}{2}(\alpha-\beta-\gamma+\alpha\beta-\beta\gamma)-\eta,\label{eq:xi_CHE}\\
\zeta	& = & \frac{1}{2}(\alpha+\beta+\gamma+\alpha\gamma+\beta\gamma)+\delta+\eta.\label{eq:zeta_CHE}
\end{eqnarray}

There are two linearly independent analytic solutions of the covariant massless Klein-Gordon equation associated with the boundaries of the effective geometry \eqref{eq:metric_UABHc}, both expressed in terms of confluent Heun functions. By deriving the asymptotic behavior of these solutions near the horizon and at spatial infinity, and subsequently imposing the QBS boundary conditions -- namely, purely ingoing waves at the horizon and decaying solutions at infinity -- it becomes possible to consistently match the solutions within their common region of validity. This procedure allows the radial equation to be reduced to a set of polynomial conditions for the confluent Heun functions. For a detailed discussion on the VBK approach, we cordially invite the readers to see the Appendix A of Ref.~\cite{PhysRevD.111.104025}.

From these polynomial conditions, namely,
\begin{eqnarray}
\frac{\delta}{\alpha}+\frac{\beta+\gamma+2}{2}+n & = & 0,\label{eq:delta-condition}\\
\Delta_{n+1}(\xi) & = & 0,\label{eq:Delta-condition}
\end{eqnarray}
one can determine the exact QBS spectrum of the effective geometry. In particular, the resulting quantization conditions provide the complete set of allowed complex frequencies characterizing the quasibound states, which are given by
\begin{equation}
\omega_{mn}	= -\frac{mC}{r_h^2}\biggl[1+\frac{a^2 r_h^4}{m^2 C^2+(n+1)^2 a^2 r_h^4}\biggr]+m\Omega-i(n+1)a\biggl[1-\frac{a^2 r_h^4}{m^2 C^2+(n+1)^2 a^2 r_h^4}\biggr].
\label{eq:omega_UABHc}
\end{equation}
where $n(=0,1,2,\ldots)$ is the overtone number. Note that the QBS frequencies satisfy the symmetry relation $\omega_{mn}=-[\omega_{-mn}]^*$, where ``*'' denotes complex conjugation. This property implies that the corresponding modes possess identical decay rates, determined by the imaginary parts of the frequencies, while their oscillation frequencies, given by the real parts, differ only by an overall sign. Consequently, the QBS spectrum is symmetric under the transformation $m \rightarrow -m$. Owing to this symmetry, it is sufficient to restrict our analysis to the corotating QBS modes only.

\begin{figure}[t]
	\centering
	\includegraphics[scale=1]{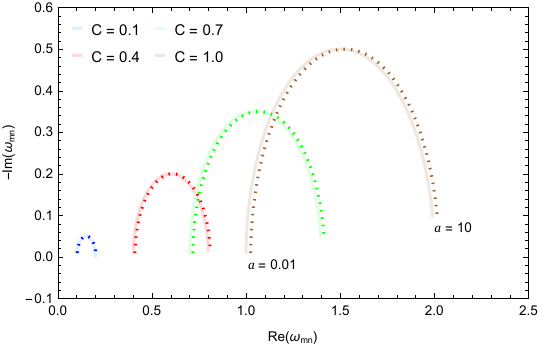}
	\caption{Fundamental ($n=0$) QBSs with unitary acoustic event horizon $r_{h}=1$, varying the tuning parameter $a$ and different choices of circulation $C$, for corotating $m=1$ modes. The solid and dotted lines corresponds to $\Omega=0$ and $\Omega=0.025C$, respectively.}
	\label{fig:Fig1_QBSs_UABHc}
\end{figure}

Figure \ref{fig:Fig1_QBSs_UABHc} displays the corotating ($m>0$) QBSs for a fixed acoustic horizon radius $r_h=1$. As discussed previously, the counterrotating ($m<0$) modes exhibit a symmetric behavior in the complex-frequency plane, sharing the same imaginary parts while differing only by the sign of their real parts. Consequently, the corresponding spectra are essentially mirror images of one another under the transformation $\mbox{Re}(\omega_{m0}) \rightarrow -\mbox{Re}(\omega_{-m0})$. This symmetry relation is preserved for all values of the circulation parameter $C$ and throughout the entire range of the tuning parameter $a$.

The observed displacement of the QBS frequencies reflects the modification of the effective trapping potential induced by the residual solid-body rotation. Since the QBS spectrum is highly sensitive to the geometry in the vicinity of the acoustic horizon, even a weak background rotation alters the balance between confinement and frame dragging, leading to measurable shifts in the complex frequency plane. This behavior indicates that the quasibound spectrum provides a sensitive probe of experimentally relevant deviations from the ideal irrotational background.

\begin{figure}[t]
	\centering
	\includegraphics[scale=1]{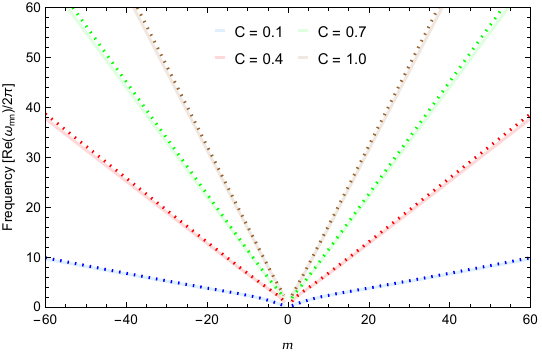}
	\caption{Two-dimensional wave spectra for the fundamental ($n=0$) QBSs with unitary tuning parameter $a=1$, unitary acoustic event horizon $r_{h}=1$ and different choices of circulation $C$. The solid and dotted lines corresponds to $\Omega=0$ and $\Omega=0.025C$, respectively.}
	\label{fig:Fig2_QBSs_UABHc}
\end{figure}

Figure \ref{fig:Fig2_QBSs_UABHc} shows the theoretical predictions for the minimum frequency required for the propagation of QBSs for a given acoustic horizon radius $r_h$. These results may be directly compared with experimental observations, such as those reported in Figs. 2(f), (g) of \cite{Nature.628.66}. In particular, the figure clearly reveals the absence of excitations within the low-frequency regime, namely below the first blue-colored line, indicating the existence of a threshold frequency for the emergence of QBS modes.

The influence of the residual solid-body rotation is cumulative rather than merely additive. Although the angular frequency $\Omega$ is assumed to be small, it modifies the global dragging profile throughout the effective spacetime, producing systematic changes in the QBS spectrum. This demonstrates that the spectral response is determined not only by the circulation parameter but also by the detailed structure of the background rotation.

\begin{figure}[t]
	\centering
	\includegraphics[scale=1]{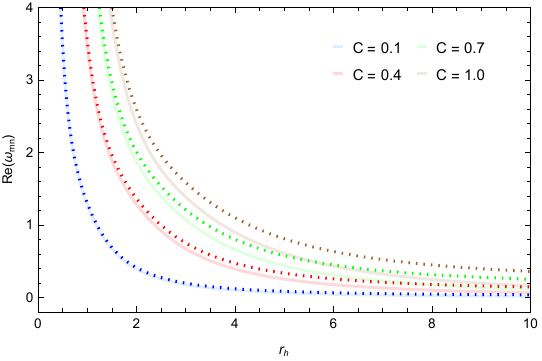}
	\caption{Oscillation frequency of the fundamental ($n=0$) QBSs with unitary tuning parameter $a=1$, varying acoustic horizon radius $r_h$ and different choices of circulation $C$, for corotating $m=8$ modes. The solid and dotted lines corresponds to $\Omega=0$ and $\Omega=0.025C$, respectively.}
	\label{fig:Fig3_QBSs_UABHc}
\end{figure}

Figure \ref{fig:Fig3_QBSs_UABHc} displays the real parts of the QBS frequencies as functions of the acoustic horizon radius of the vortex. For a fixed tuning parameter $a=1$, the figure reveals a remarkably structured dependence of the QBS spectrum on the radius of the rotating acoustic black hole, $r_h$. In particular, the parabolic profiles exhibited by the QBS frequencies for different values of the circulation parameter $C$ become increasingly pronounced as the acoustic horizon grows, especially for the corotating mode with $m=8$. Interestingly, similar parabolic patterns have recently been observed experimentally in a giant hollow-core quantum vortex of superfluid $^4$He \cite{Nature.628.66}. Although our results qualitatively resemble those experimental findings, no direct correspondence can yet be established, mainly because the superfluid vortex analyzed in \cite{Nature.628.66} does not possess radial flow and, consequently, lacks an acoustic horizon. Therefore, the present results describe a more intricate configuration and may provide further insights for future theoretical and experimental investigations.

The approximately hyperbolic behavior of the QBS frequencies reflects the competition between two characteristic length scales of the problem: the acoustic horizon radius and the effective rotational length associated with the circulating flow. As the horizon radius increases, the effective potential well broadens and the localization properties of the QBSs change accordingly, giving rise to the observed nonlinear dependence of the spectrum. Equation (\ref{eq:omega_UABHc}) explains the hyperbolas that we observe in Fig. \ref{fig:Fig3_QBSs_UABHc}

\begin{figure}[t]
	\centering
	\includegraphics[scale=1]{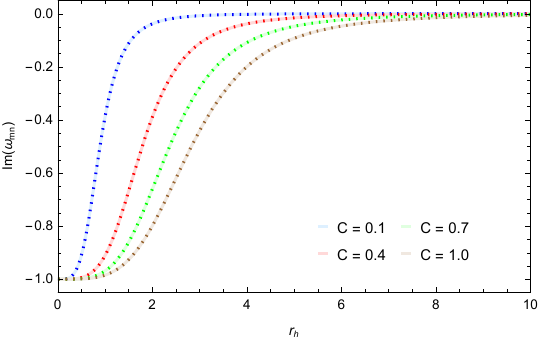}
	\caption{Decay rate of the fundamental ($n=0$) QBSs with unitary tuning parameter $a=1$, varying acoustic horizon radius $r_h$ and different choices of circulation $C$, for corotating $m=8$ modes. The solid and dotted lines corresponds to $\Omega=0$ and $\Omega=0.025C$, respectively.}
	\label{fig:Fig4_QBSs_UABHc}
\end{figure}

In finality, Fig. \ref{fig:Fig4_QBSs_UABHc} displays the imaginary parts of the QBS frequencies as functions of the acoustic horizon radius of the vortex.

The circulation parameter primarily controls the rotational contribution inherited from the vortex flow, whereas the solid-body component represents a background correction associated with the experimental apparatus. Their combined action determines the effective frame-dragging profile, explaining why different values of the circulation parameter respond differently to the same residual rotation.

Taken together, these results suggest that QBSs constitute particularly sensitive observables for detecting small departures from idealized acoustic geometries. Since the residual solid-body rotation modifies the spectrum in a systematic manner, future analog gravity experiments may use spectral measurements not only to identify QBSs but also to quantify background rotational effects that are usually neglected in ideal theoretical models.
%
%%%%%%%%%%%%%%%%%%%%%%%%%%%%%%%%%%%%%%%%% Conclusions
%
\section{Conclusions}\label{Conclusions}
%
%%%%%%%%%%%%%%%%%%%%%%%%%%%%%%%%%%%%%%%%%
%
The present work introduces a phenomenological extension of the standard acoustic geometry that incorporates experimentally motivated residual rotation. Although the model is not intended as an exact solution of the hydrodynamic equations, it provides an analytically tractable framework for investigating how background rotation affects quasibound state spectra.

Analog gravity has become one of the most powerful approaches for investigating strong-gravity phenomena in controlled laboratory settings, encompassing a broad range of experimental platforms based on distinct physical mechanisms, including water waves, optical systems, lasers, condensates, electronic circuits, and several others. These analog systems provide an important framework for exploring quantum effects in the vicinity of black holes, as well as classical gravitational phenomena that remain beyond the reach of current observational capabilities. In this work, we constructed an effective geometry describing a rigidly rotating superfluid with circulation and a residual solid-body rotation in the background flow, yielding a spacetime structure that closely resembles that of a rotating black hole while remaining free of singular regions. By analyzing the propagation of surface acoustic waves in this effective geometry, we derived the corresponding QBS spectrum from a theoretical perspective. Our results exhibit qualitative agreement with previous experimental observations reported in Refs.~\cite{Nature.628.66,PhysRevLett.121.061101,PhysRevLett.125.011301}.

The analysis of the QBSs appears to support, at least qualitatively, the findings reported in Ref.~\cite{Nature.628.66}, although within a different and potentially promising theoretical framework. In particular, the QBS spectrum was obtained for both corotating and counterrotating acoustic perturbations propagating in our effective rotating black hole geometry.

The future prospects of such analog-gravity experiments are extensive and may help clarify several phenomenological aspects of black hole physics at both the quantum and classical levels. In particular, a highly interesting direction concerns the implementation of environmental effects in analog black hole systems. The inclusion and investigation of astrophysical environments surrounding black holes currently constitute one of the most active and promising areas in gravitational-wave astrophysics. In this context, it would be especially interesting to extend experiments such as the one reported in Ref.~\cite{Nature.628.66} in order to model a hollow-core vortex with circulation and an acoustic horizon, supplemented either by external impurities that do not destabilize the vortex configuration or by nonlinear dispersive interactions within a single superfluid capable of modifying the radial velocity profile of the flow. Such configurations could effectively emulate analog black holes surrounded by accretion disks, black hole hair \cite{PhysRevD.103.044042,GenRelGrav.54.49}, or even dark matter halos around supermassive black holes at galactic centers \cite{PhysRevD.105.L061501,PhysRevD.107.084027}.

The information extracted from such realistic analog systems may play an important role in improving our understanding of strong-field gravity and the behavior of matter in the vicinity of black holes \cite{PhysRevD.111.024017,PhysRevD.111.064026}. In addition, the spatial profiles of the eigensolutions obtained in the present work may provide valuable guidance for future experimental investigations involving dissipative effects in analog black hole configurations.

Our analysis further indicates that the inclusion of a residual solid-body rotation in the background flow introduces subtle, yet non-negligible, modifications to the QBS spectrum of the effective acoustic black hole geometry. Although these corrections remain relatively small, they systematically affect the structure of the QBS frequencies and their dependence on the acoustic horizon and circulation parameters, potentially leading to observable signatures in future high-precision experiments. In this sense, the present model provides a more realistic and refined description of rotating superfluid vortices with acoustic horizons, extending previous analog black hole frameworks. Therefore, the effective geometry introduced here may serve as a useful theoretical platform for confronting and interpreting forthcoming experimental observations, as well as for testing the influence of residual rotational effects on analog strong-gravity phenomena.

More generally, the present construction illustrates how experimentally motivated departures from ideal irrotational backgrounds can be incorporated into analytically solvable analog spacetimes. We expect that similar effective extensions may prove useful for investigating other environmental effects in analogue black-hole systems.
%
%%%%%%%%%%%%%%%%%%%%%%%%%%%%%%%%%%%%%%%%% Acknowledgments
%
\begin{acknowledgments}
%
%%%%%%%%%%%%%%%%%%%%%%%%%%%%%%%%%%%%%%%%%
%
This study was financed in part by the Conselho Nacional de Desenvolvimento Científico e Tecnológico -- CNPq -- Research Project No. 446211/2024-9 and Research Fellowship No. 314926/2025-9.
%
%%%%%%%%%%%%%%%%%%%%%%%%%%%%%%%%%%%%%%%%%
%
\end{acknowledgments}
%
%%%%%%%%%%%%%%%%%%%%%%%%%%%%%%%%%%%%%%%%% References
%

%
%%%%%%%%%%%%%%%%%%%%%%%%%%%%%%%%%%%%%%%%%%%%%%%%%%%%%%%%%%%%%%%%%%%%%%%%%%%%%%%%%%%%%%%%%%%%%%
%
\end{document}